\documentclass[preprint]{aastex}

\newcommand{\wisk}[1]{\ifmmode{#1}\else{$#1$}\fi}

\newcommand{\bfgamma}{\mbox{\boldmath $\gamma$}}
\newcommand{\bfomega}{\mbox{\boldmath $\omega$}}

\def\gsim{\;_\sim^>\;}
\def\lsim{\;_\sim^<\;}
\def\half{\frac{1}{2}}
\slugcomment{IUCAA-22/2001, \hspace{0.2truecm} KSUPT-01/3 \hspace{0.5truecm} May 2001}

\begin{document}

\title{Window Function for Non-Circular Beam CMB Anisotropy Experiment}

\author{Tarun Souradeep\altaffilmark{1,2} 
        and 
        Bharat Ratra\altaffilmark{1}}

\altaffiltext{1}{Department of Physics, Kansas State University, Manhattan, 
                 KS 66506.}
\altaffiltext{2}{Current address: IUCAA, Post Bag 4, Ganeshkhind, Pune 411007,  
                 India.}

\begin{abstract}
We develop computationally rapid methods to compute the window function
for a cosmic microwave background anisotropy experiment with a
non-circular beam which scans over large angles on the sky. To concretely 
illustrate these methods we compute the window function for the Python V 
experiment which scans over large angles on the sky with an elliptical 
Gaussian beam.  
\end{abstract}


\section{Introduction}

Cosmic microwave background (CMB) anisotropy measurements are becoming
an increasingly powerful tool for testing cosmogonies and constraining
cosmological parameters. See, e.g., Subrahmanyan et al. (2000), Romeo et
al. (2001), Dawson et al. (2000), and Padin et al. (2001) for recent CMB
anisotropy observations, and, e.g., Ratra et al. (1997), G\'orski et al. 
(1998), Rocha et al. (1999), Gawiser \& Silk (2000), Knox \& Page (2000), 
Douspis et al. (2001), and Podariu et al. (2001) for discussions of 
constraints on models from the CMB anisotropy data.

Conventionally, the CMB temperature, $T(\bfgamma)$, is expressed as a function
of angular position, $\bfgamma\equiv(\theta, \phi)$, on the sky via
the spherical harmonic decomposition,
\begin{equation}
   T(\bfgamma) = \sum^\infty_{\ell = 0}
   \sum^\ell_{m = -\ell} a_{\ell m} Y_{\ell m} (\bfgamma)\,.
   \label{dT_ylm}
\end{equation}
The CMB spatial anisotropy in a Gaussian model\footnote{ 
  The simplest inflation models predict a Gaussian CMB anisotropy (see, e.g.,
  Fischler, Ratra, \& Susskind 1985) on all but the smallest angular
  scales. CMB anisotropy observations on quarter degree and larger angular
  scales appear to be Gaussian (see, e.g., Mukherjee, Hobson, \&
  Lasenby 2000; Aghanim, Forni, \& Bouchet 2001; Phillips \& Kogut
  2001; Park et al. 2001; Wu et al. 2001b, also see Podariu et al. 2001).} 
is completely specified by its angular two-point correlation function
$C(\bfgamma,\bfgamma^\prime)$, $= \langle T(\bfgamma)
T(\bfgamma^\prime)\rangle$, between directions $\bfgamma$ and
$\bfgamma^\prime$ on the sky. In most theoretical models the
predicted fluctuations are statistically isotropic, $
C(\bfgamma,\bfgamma^\prime) \equiv C(\bfgamma\cdot\bfgamma^\prime)$.
The fluctuations can then be characterized solely by the angular
spectrum $C_\ell$, defined in terms of the ensemble average,
\begin{equation}
   \langle a_{\ell m} a_{\ell^\prime m^\prime}{}^* \rangle =
   C_\ell \,\delta_{\ell\ell^\prime} \delta_{mm^\prime}\,,
   \label{alm_Cl}
\end{equation}
and  related to the correlation function through
\begin{equation}
   \langle T(\bfgamma)\,T(\bfgamma^\prime)\rangle =\sum^\infty_{\ell = 0} 
   {(2\ell + 1) \over 4\pi} \,C_\ell \,  P_\ell(\bfgamma\cdot\bfgamma^\prime)\,,
   \label{corr_Cl}
\end{equation}
where $P_\ell$ is a Legendre polynomial.

Typically, a CMB anisotropy experiment probes a range of angular scales 
characterized by a window function $W_\ell(\bfgamma, 
\bfgamma^\prime)$.\footnote{
  See www.phys.ksu.edu/$\sim$tarun/CMBwindows/wincomb/wincomb\_tf.html
  for a discussion and tabulation of zero-lag window functions.}
To utilize the full information in the data one must use model anisotropy 
spectra ($C_\ell$'s) defined over this range of angular scales. Such 
theoretical spectra are parameterized by cosmological parameters such as 
$\Omega_0$, $h$, and $\Omega_B$ in these models\footnote{ 
  Here $\Omega_0$ is the nonrelativistic-mass density
  parameter, $h$ is the Hubble parameter in units of $100\ {\rm km}\ 
  {\rm s}^{-1}\ {\rm Mpc}^{-1}$, and $\Omega_B$ is the baryonic-mass
  density parameter.}, 
and by the spectrum of quantum fluctuations generated during inflation.

To use model $C_\ell$'s in conjunction with CMB anisotropy data to
estimate cosmological parameters one must be able to carefully model
the CMB anisotropy experiment, i.e., accurately compute the window
function $W_\ell$. Given such a model of an experiment and a family of
$C_\ell$'s, one may optimize the fit to the data from the experiment
either by using an (approximate) $\chi^2$ technique (e.g., Ganga,
Ratra, \& Sugiyama 1996; Bond, Jaffe, \& Knox 2000; Knox 1999; 
Rocha 1999; Lineweaver 2001; Dodelson 2000;
Tegmark \& Zaldarriaga 2000) or by using an exact maximum likelihood
technique (e.g.,  G\'orski et al. 1995; Ganga et al. 1997, 1998; 
Ratra et al. 1998, 1999).

Current CMB anisotropy data is of significantly higher quality than
data that was available just a few years ago. As a consequence an
accurate model of an experimental $W_\ell$ must now account for
effects that were ignored for earlier experiments. In this paper we
develop computationally rapid methods that account for the
non-circularity of the beam in a CMB anisotropy experiment window
function at large angular separations where the curvature of the sky
cannot be ignored.  This must be accounted for in an experiment like
Python V (Coble et al.  1999) which has an elliptical beam and samples
a large enough area of the sky to prejudice use of the flat-sky
approximation.  While we focus here on Python V as a concrete
illustrative example, our techniques are easily generalized to more
complex cases (e.g., arbitrary beam shape, beam rotation,
non-Gaussianity of the beam, etc.). Wu et al. (2001a) develop an alternate
method to deal with the asymmetric beam of the MAXIMA-1 experiment. 
 
In $\S$\ref{wfsec} we describe the general formalism for computing the
window function. In the $\S\S$\ref{wf_flat}--\ref{wf_eqdec} we develop 
specific computationally rapid methods for computing the window function 
in three different cases. The flat-sky approximation window function 
computation is covered in $\S$\ref{wf_flat}. In $\S$\ref{wf_wig} we develop 
a general method, based on Wigner rotation functions, for computing the 
window function on a sphere, and describe how to numerically implement this 
scheme. In $\S$\ref{wf_eqdec} we specialize to the case of an experiment 
like Python V where long scans are performed at constant elevation, and 
provide a computationally rapid method for evaluating the exact window 
function. Approximate window functions obtained using the flat-sky 
approximation and from retaining only the first few terms in a perturbation
expansion (in non-circularity about a circular beam) of the Wigner
rotation functions method (hereafter Wigner method) are compared to
the exact window function in $\S$\ref{wincomp}. We conclude in
$\S$\ref{concl}. In appendix~\ref{analbeam} we describe our parameterization 
and normalization of an elliptical Gaussian beam, and also record analytic 
expressions for its Fourier and spherical harmonic transforms.

\section{ Window Function Computation Formalism}
\label{wfsec}

Owing to the finite angular resolution of a CMB anisotropy experiment, the 
temperature ``measured" by the experiment at point ${\bfgamma}_i$ on the sky 
is
\begin{equation}
   \tilde T({\bfgamma}_i) = \int d\Omega_{\bfgamma}\,
   B\left({\bfgamma}_i,{\bfgamma}\right) T({\bfgamma}) .
   \label{dTbeam}
\end{equation}
Here $B({\bfgamma}_i, {\bfgamma})$ is the beam function that characterizes 
the angular dependence of the sensitivity of the apparatus around the 
pointing direction ${\bfgamma}_i$.

CMB anisotropy experiments that use a differencing or modulation scheme 
measure the difference in temperature between different points on the 
sky. The measured CMB temperature anisotropy in any differencing scheme 
(labelled below by index $n$) can be expressed as a weighted linear 
combination,
\begin{equation}
   \Delta^{(n)}({\bfgamma}_i) = \int d\Omega_{\bfgamma}\, w^{(n)}_i
   (\bfgamma)\,\,\tilde T({\bfgamma}) \,,
   \label{wtcombdT}
\end{equation}
where $w^{(n)}_i(\bfgamma)$ are the weight functions. In our discussion 
of Python V below $n$ corresponds to the harmonic number of cosine 
modulation weight functions. Instead of weights $w^{(n)}_i(\bfgamma)$ that 
are continuous functions, for Python V the weights $w^{(n)}_{ij}$ are 
discrete and characterize the sensitivity at $\bfgamma_{ij}$, the 
$j^{\rm th}$ point on the discretized chopper cycle around the pointing 
direction $\bfgamma_i$. In this case the integral in eq. (\ref{wtcombdT}) 
is replaced by a summation and we have
\begin{equation}
   \Delta^{(n)}({\bfgamma}_i) =\sum_{j=1}^{N_c} w^{(n)}_{ij}
   \,\tilde T({\bfgamma_{ij}})
   \label{wtcombtemp}
\end{equation}
where $N_c$ is the number of points in the discretized chopper cycle.

The complete window function for modulation pair $(n,m)$,
$W_\ell^{(n,m)}(\bfgamma_i,\,\bfgamma_j)$, accounts for the effects of both 
the beam function and the differencing or modulation scheme of the 
experiment and is defined through the theoretical model covariance matrix
\begin{equation}
   C^{(n,m)}_{ij} = \langle \Delta^{(n)}({\bfgamma}_i) 
   [\Delta^{(m)}({\bfgamma}_j)]^* \rangle
   = \sum^\infty_{\ell = 0} {(2\ell + 1) \over 4\pi}\,\, C_\ell\,\,
   W_\ell^{(n,m)}(\bfgamma_i,\, \bfgamma_j)\,.
   \label{ctij_wl}
\end{equation}
It proves convenient to distinguish between the dependence of the window 
function on the finite angular resolution of the experimental apparatus 
(the beam function) and the dependence of the window function on the differencing scheme adopted for the experiment. Using eq. (\ref{wtcombtemp}), 
the complete theoretical covariance matrix element $C^{(n,m)}_{ij}$ between 
pixels $i$ and $j$ on the sky can be expressed as a weighted linear sum of 
single beam correlation functions via
\begin{equation}
   C^{(n,m)}_{ij}\equiv  \sum^{N_c}_{p=1} \sum^{N_c}_{q=1} 
   w^{(n)}_{i p}\, w^{(m)}_{jq}
   \,\,\langle \tilde T(\bfgamma_{ip}) \tilde T(\bfgamma_{jq})\rangle\,.
   \label{ctij_corre}
\end{equation}
We call the single beam correlation function $C^{({\rm e})} 
(\bfgamma,\bfgamma^\prime)$, $= \langle \tilde T({\bfgamma})
\tilde T({\bfgamma}^\prime)\rangle$, an elementary correlation function.
The elementary correlation function does not depend on the differencing 
scheme used in the experiment but does depend on the beam function.
We use $C^{({\rm e})}$ to define what we call the (single beam) elementary 
window function $W_\ell^{({\rm e})}$ via
\begin{equation}
   C^{({\rm e})}(\bfgamma_i,\bfgamma_j)   =
   \sum^\infty_{\ell = 0} {(2\ell + 1) \over 4\pi}\, C_\ell\,\,
   W_\ell^{({\rm e})}(\bfgamma_i,\, \bfgamma_j)\,.
   \label{corre}
\end{equation}
$W_\ell^{({\rm e})}$ depends on the beam function of the experiment but not 
on the differencing strategy used. Using eqs.~(\ref{ctij_corre}) and
(\ref{corre}) the complete window function may be expressed as a weighted
linear sum of elementary window functions via
\begin{equation}
   W_\ell^{(n,m)}(\bfgamma_i,\, \bfgamma_j) = \sum^{N_c}_{p=1}
   \sum^{N_c}_{q=1} w^{(n)}_{i p}\, w^{(m)}_{jq}\,\,  W_\ell^{({\rm
   e})}\left(\bfgamma_{i p},\, \bfgamma_{j q}\right).
   \label{win_wine}
\end{equation}

Using eqs.~(\ref{corr_Cl}), (\ref{dTbeam}), and (\ref{corre}) the elementary 
window function may be expressed as
\begin{eqnarray}
   W_\ell^{({\rm e})}(\bfgamma_i,\, \bfgamma_j)
   & = & \int
      d\Omega_{\bfgamma}\, d\Omega_{\bfgamma^\prime}
      B(\bfgamma_i, \, \bfgamma) B(\bfgamma_j, \,
      \bfgamma^\prime) P_\ell(\bfgamma\cdot\bfgamma^\prime)\, 
      \nonumber \\ 
       & = & \frac{4\pi}{2\ell +1}\,\,
     \sum_{m=-\ell}^\ell \, b_{\ell m}({\bfgamma}_i)\, [b_{\ell
      m}({\bfgamma}_j)]^*,
\label{genwine}
\end{eqnarray}
where
\begin{equation}
    b_{\ell m} ({\bfgamma}_i) = \int d\Omega_{\bfgamma}\ B\left({\bfgamma}_i,
    {\bfgamma}\right) Y^*_{\ell m}(\bfgamma)
    \label{blm}
\end{equation}
is the spherical harmonic transform of the beam function pointing at
$\bfgamma_i$.

For some experiments the beam function is accurately circularly symmetric about
the pointing direction, i.e., $B({\bfgamma}_i, {\bfgamma}) \equiv
B({\bfgamma}_i\cdot {\bfgamma})$.  This allows a great
simplification\footnote{ When the beam is pointing towards a Pole,
  $\bfgamma_{\rm P}$, the coefficients in the spherical harmonic
  expansion of a circular beam function are $b_{\ell m} (\bfgamma_{\rm
    P}) = B_\ell \sqrt{(2\ell +1)/4\pi}\,\delta_{m 0}$. Here
  $\delta_{m 0}$ is a Kronecker delta function and implies that the
  only non-zero spherical harmonic coefficients are those that result
  in a circularly-symmetric beam function.}  since the beam function
can then be represented as $B({\bfgamma}_i\cdot {\bfgamma}) =
(4\pi)^{-1}\,\sum_{\ell=0}^\infty\, (2\ell+1)\, B_\ell\,
P_\ell({\bfgamma}_i\cdot {\bfgamma})$. Consequently, for a circularly
symmetric beam function it is straightforward to derive the usual
expression
\begin{equation}
   W_\ell^{({\rm e})}(\bfgamma_i,\, \bfgamma_j) = B^2_\ell \,
   P_\ell({\bfgamma}_i\cdot {\bfgamma}_j)\,.
   \label{isowine}
\end{equation}
In addition to the single evaluation of the Legendre transform of the
beam, $B_\ell$, computation of the elementary window function for an
experiment with a circular beam simply involves computing $P_\ell$ for
all $\ell \le \ell_{\rm max}$ using the stable upward recursion
relation for each distinct pixel pair separation. This is
computationally inexpensive (at least by a factor $\sim\ell_{\rm max}$)
compared to the computation of the elementary window function for an 
experiment with arbitrary beam shape (eq. [\ref{genwine}]).

In the next three sections we discuss three cases where one may compute 
the window function for an experiment with a non-circular beam in a time 
comparable to or just a factor of few larger than that for the same 
experiment assuming  a circular beam. First, we consider the flat-sky
approximation which is accurate if the experiment has a compact beam and 
the pixels are not separated by large angles (more precisely, the separation 
must be significantly less than a radian). This has been used previously
for a number of experiments,  including Python V (Coble et al. 1999; Coble 
1999), MSAM (Coble 1999), and MAXIMA-1 (Wu et al. 2001a). Next, we develop a 
very general Wigner method that fully 
accounts for the curvature of the sky and accounts for the non-circularity 
of the beam in a perturbative expansion about a circular beam. In a case 
like Python V where the non-circularity of the beam is not large, the first 
few terms in the non-circularity perturbation expansion provide
sufficient accuracy. In this case the Wigner method allows one to compute 
the window function in a time a factor of a few larger than that for the 
corresponding circular beam case. Finally, for an experiment that scans 
at constant elevation (such as Python V), with pixels lying on a small 
number of elevations, it is possible to implement a slightly different 
Wigner method that allows rapid computation of the exact window function 
for an arbitrary beam shape.

\section{Window Function in the Flat-Sky Approximation}
\label{wf_flat}

If an experiment scans a small enough patch of the sky it is
computationally advantageous to work in the flat tangent plane (rather
than on the sphere) and make use of Fourier transforms (rather than
spherical harmonic transforms) when modeling the experiment. See, for
instance, Bond \& Efstathiou (1987) and Coble (1999) for discussions. 
We may then transform from ${\bfgamma}$ to coordinates in a locally 
flat patch, ${\bfomega}$, and use a two-dimensional Fourier transform
approximation to the spherical harmonic transform. For instance,
\begin{equation}
   T({\bfomega}) = \int {d^2k \over (2\pi)^2}\, e^{i {\bf k} \cdot {\bfomega}}
   T({\bf k}).
   \label{dT_ft}
\end{equation}
Here ${\bfomega},\, = (\omega ,\, \phi)$, are polar coordinates in the
neighborhood of the North Pole ${\bfgamma}_{\rm P}$ in the patch on
the sky, i.e., ${\bfgamma} = {\bfgamma}_{\rm p} + {\bfomega}$, or
${\bfomega} = (\omega_1,\, \omega_2) = (\omega \cos\phi,\,\omega
\sin\phi)$, where $\omega = 2\,\sin(\theta/2) = |{\bfgamma} -
{\bfgamma}_{\rm P}|$, $0 \leq \omega \leq 2$, where $\theta$ is the
colatitude (e.g., Bond \& Efstathiou 1987).  In the small-angle
approximation, it may be shown that the ensemble average of the
Fourier transform of the temperature 
\begin{equation}
   \langle T({\bf k}) T^*({\bf k}^\prime) \rangle = (2\pi)^2
   C_k \delta^{(2)} ({\bf k} - {\bf k}^\prime)\,,
   \label{dTk_Ck}
\end{equation}
where $k=|{\bf k}|$ and $C_k$ is the CMB anisotropy power spectrum. This 
flat-sky analog of eq.~(\ref{alm_Cl}) is
obtained assuming statistical homogeneity on the flat sky. The angular
correlation function can then be expressed as
\begin{equation}
   \langle T({\bfomega_i})\,T({\bfomega_j})\rangle = \frac{1}{2\pi} 
   \,\, \int_0^\infty dk\, k\, J_0(k|\bfomega_i-\bfomega_j|)\,C_k\,,
   \label{fcorr_Ck}
\end{equation}
where $J_0$ is the zeroth order Bessel function of the first kind. Comparing 
this expression to that in eq.~(\ref{corr_Cl}) in the small angular 
separation and large $\ell$ regime where $P_\ell(\cos\theta)\to 
J_0([\ell+1/2]\theta)$, we arrive at the correspondence $k \sim \ell+1/2$ 
between the radial wavenumber on the flat sky $k$ and the spherical 
multipole $\ell$.  

For an experiment for which the flat-sky approximation is valid the 
``measured'' temperature (see eq.~[\ref{dTbeam}]) is
\begin{equation}
   \tilde T(\bfomega_i) =  \int {d^2k \over (2\pi)^2}\,\, e^{i {\bf k}
   \cdot {\bfomega}_i}\,\, B({\bf R}_i[{\bf k}]) \,T({\bf k}) ,
   \label{fldT}
\end{equation}
where $B({\bf k})$ is the Fourier transform of the beam function
pointing at the origin $\bfomega = 0$ of the local flat coordinate
patch. The rotation operator ${\bf R}_i$ (which rotates ${\bf k}$ by
an angle $\varrho_i$) accounts for a possible rotation of the
telescope beam pointing at $\bfomega_i$ relative to the telescope beam
pointing at the origin. In addition to the case when the telescope
physically rotates around its axis as it moves from one pointing
direction to another, non-zero values of $\varrho_i$ can arise when
the telescope is not located at a Pole and also when a single flat-sky
coordinate system is set up on a patch large enough for sky curvature
to be important.  In the latter two cases this rotation is important
only in the regime where one expects the flat-sky approximation to be
poor.

The modulated temperature $\Delta^{(n)}({\bfgamma}_i)$ expressed in
terms of $\tilde T(\bfomega_{ij})$ is given by eq.~(\ref{wtcombtemp}).
The window function $W_\ell^{(n,m)}(\bfgamma_i,\, \bfgamma_j)$ is
defined through the covariance matrix by
\begin{equation}
   C^{(n,m)}_{ij} = \sum^\infty_{\ell = 0} {(2 \ell + 1) \over 4\pi}\,\,
   C_\ell \,W_\ell^{(n,m)}(\bfgamma_i,\,\bfgamma_j) \simeq
   \int^\infty_0 {dk \over 2\pi}\, k\, C_k\,
   W_k^{(n,m)}(\bfgamma_i,\,\bfgamma_j)\,,
   \label{ctij_Wk} 
\end{equation}
where we identify the flat space radial wavenumber $k$ with $\ell +
1/2$. The complete window function $W_k^{(n,m)} (\bfgamma_i,\,
\bfgamma_j)$ may be expressed in terms of elementary window functions
$W_k^{({\rm e})}$ by using eq.~(\ref{win_wine}).

Using eqs.~(\ref{corre}) and (\ref{fldT}), the expression for the
elementary window function in the flat-sky approximation is
\begin{equation}
   W_k^{({\rm e})}(\bfomega_i,\, \bfomega_j)
   = \int^{2\pi}_0 {d\phi_k \over 2\pi}\, e^{i {\bf k} \cdot ({\bfomega}_i - 
   {\bfomega}_j)} B({\bf R}_i[{\bf k}]) B^*({\bf R}_j[{\bf k}])\,.
   \label{flatwine}
\end{equation}
Further analytical manipulations are needed to derive an expression suitable 
for numerical evaluation. Without loss of generality, we transform to a new 
flat coordinate system with $\bfomega_i$ as the origin: $\bfomega^\prime_i = 
(0,0)$ and $\bfomega^\prime_j = (\omega^\prime_1, \omega^\prime_2)$. In what 
follows we drop the prime on the new coordinates. For small angular 
separations, $\omega = |\bfomega_i-\bfomega_j|\ll 1$, we have
\begin{equation}
   W_k^{({\rm e})}(\bfomega_i,\, \bfomega_j) 
   = \int^{2\pi}_0 {d\phi_k \over 2\pi}\,
   {\rm cos}\left[k \,\omega \, {\rm cos}(\phi_k - \alpha)\right]
   \,B({\bf k})\,B^*({\bf R}_j[{\bf k}])\,,
   \label{flatwine2}
\end{equation}
where we have defined $\alpha = {\rm tan}^{-1}(\omega_2/\omega_1)$.

For an elliptical Gaussian beam function (eq.~[\ref{angau_beam}]), an analytic expression for $B({\bf k})$ is given in eq.~(\ref{pyv_beam_ft}). Using this in
eq.~(\ref{flatwine2}) (and ignoring rotations, $\varrho_j = 0$) we obtain
\begin{equation}
   W_k^{({\rm e})}(\bfomega_i,\, \bfomega_j) 
   = \int^{\pi}_0 {d\phi_k \over \pi}\,\,
   {\rm cos}\left[k\, \omega \,{\rm cos}(\phi_k - \alpha)\right]\,\,
   \exp\left[-k^2 \sigma_1{}^2 (1 + \varepsilon\, {\rm sin}^2 \phi_k)\right]\,,
   \label{flatwine_pyv}
\end{equation}
where $\sigma_1$ and $\sigma_2$ are the beamwidths (in radians) along
the major and minor axis of the elliptical Gaussian beam. It is useful
to define the RMS beamwidth $\sigma_{\rm RMS}$, $= \sqrt{(\sigma_1{}^2
  + \sigma_2{}^2)/2}$, and a measure of the non-circularity of the
elliptical beam $\varepsilon$, $= (\sigma_2/\sigma_1)^2- 1$.

The expression for the flat-sky window function for an elliptical Gaussian
beam given in eq.~(\ref{flatwine_pyv}) can be readily evaluated numerically.
It can also be expressed analytically as an infinite series expansion,
$ W_k^{({\rm e})} = \sum_{n=0}^\infty (-1)^n \, b_k^n\,\,
{}^{(n)}W_k^{({\rm e})}$, in powers of an anisotropy or non-circularity 
parameter $b_k = \varepsilon(\sigma_1 k)^2 = k^2 (\sigma_2^2 -\sigma_1^2)$. The
series is perturbative ($b_k < 1$) up to multipoles $\ell \lsim
|\sigma_2^2 -\sigma_1^2|^{-1/2}$, which can be considerably larger than
the inverse of the RMS beamwidth $\sigma_{\rm RMS}$. As expected, the first 
($n=0$) term in this series corresponds to the circular beam function result 
($\sigma_1=\sigma_2=\sigma$),
\begin{equation}
   {}^{(0)}W_k^{({\rm e})}(\bfomega_i,\, \bfomega_j) = J_0\left( k
   \omega \right) \, e^{-k^2 \sigma_1{}^2}\,.
   \label{wf0}
\end{equation}
For a non-circular Gaussian beam function the next few terms are
\begin{eqnarray}
   && {}^{(1)}W_k^{({\rm e})}(\bfomega_i,\, \bfomega_j) = \left[
      \frac{J_1( k \omega)}{k \omega} - J_2( k \omega)\,
      \sin^2\alpha \right] \, e^{-k^2 \sigma_1{}^2}  
\label{wf_1}\\
   && {}^{(2)}W_k^{({\rm e})}(\bfomega_i,\, \bfomega_j) =
      \frac{1}{2}\,\left[ 3 \frac{J_2( k \omega)}{(k \omega)^2} - 6
      \frac{ J_3( k \omega)}{k \omega}\, \sin^2\alpha + J_4(k
      \omega)\sin^4\alpha \right]\,e^{-k^2 \sigma_1{}^2} \,,
   \label{wf_2}
\end{eqnarray}
where $J_m(x)$ is the $m^{\rm th}$ order Bessel function of the first kind.
At arbitrary order $n$ the term is
\begin{eqnarray}
   {}^{(n)}W_k^{({\rm e})}(\bfomega_i,\, \bfomega_j) =&& 
   \frac{\sqrt{\pi}\,\Gamma(n+1/2)}{\Gamma(n+1)} e^{-k^2
   \sigma_1{}^2} \sum_{m=0}^n
   \frac{(\sin\alpha)^{2 m}
   (\cos\alpha)^{2(n-m)}}{\Gamma(n-m+1)\,\Gamma(m+1)} \nonumber\\ &&{}
   \times \,\,{}_1\!F_2\left[\half+m,\left( \half, 1 + n \right),
   -\frac{(k \,\omega)^2}{4} \right] \,,
   \label{wf_n}
\end{eqnarray}
where $\Gamma$ is the Euler gamma function and ${}_1F_2$ is a generalized
hypergeometric function. Figure~1 shows contour plots of the
zeroth order (eq.~[\ref{wf0}]) and first order (eq.~[\ref{wf_1}]) terms
in the non-circularity expansion of the flat-sky window function. These are
computed for parameter values characterizing the Python V experiment (see 
$\S$\ref{wincomp} below for details of the experiment).

\section{Wigner Method Window Function}
\label{wf_wig}

If an experiment takes data over a large enough area of the sky the
formalism developed in the previous section, based on the flat-sky
approximation, cannot be used to compute the window function. In this
section we develop a general method for computing the window function
for an experiment with an arbitrary beam shape that collects data from
a large area on the sky. 

For pointing direction $\bfgamma_i$, $= (\theta_i,\, \phi_i)$, and vector 
in the beam $\bfgamma$, $= (\theta,\, \phi)$,  the beam function may be 
expanded in a spherical harmonic decomposition,
\begin{equation}
   B(\bfgamma_i ,\, \bfgamma) = \sum^\infty_{\ell = 0}
   \sum^\ell_{m = -\ell} b_{\ell m}(\bfgamma_i) Y_{\ell m} (\bfgamma)\,.
   \label{beam_ylm}
\end{equation}
Here the expansion coefficients, $b_{\ell m}$, are given by
eq.~(\ref{blm}).

For ease of computation it is convenient to rotate to a new coordinate
system in which the new ${\bf x}_3^\prime$ axis lies along the
pointing direction $\bfgamma_i$. This is accomplished by first
rotating the coordinate system around the ${\bf x}_3$ axis by $\phi_i$
and then rotating around the new ${\bf x}_2^\prime$ axis by
$\theta_i$. Then
\begin{equation}
   B({\bf x}_3^\prime ,\, \bfgamma^\prime) = \sum^\infty_{\ell^\prime = 0}\, 
   \sum^{\ell^\prime}_{m^\prime = -\ell^\prime} b_{\ell^\prime m^\prime}
   ({\bf x}_3^\prime) Y_{\ell^\prime m^\prime} (\bfgamma^\prime)
   =  B(\bfgamma_i ,\, \bfgamma)\,,
   \label{beamrot}
\end{equation}
where the last step follows from the fact that $B$ is a scalar.  

If the experiment is not located at a Pole, the beam of a telescope which does 
not rotate around its beam axis will, nevertheless, appear to ``rotate'' around 
the beam axis with respect to the local azimuth and declination directions.  
We hence allow for a rotation $\varrho_i$ of the beam around the beam axis 
relative to a parallel transport of the beam on the sky from the Pole 
$\bfgamma_{\rm P}^\prime$,  $= {\bf x}_3^\prime$, to the pointing direction 
$\bfgamma_i^\prime$. The Python V experiment was located at the South Pole 
and hence has $\varrho_i=0$. The rotation $\varrho_i$ can also account for 
non-circular beam function cases where the telescope rotates around its 
axis as it moves from one pointing direction to another (e.g., one mounted on 
a satellite).

The rotations of the previous two paragraphs correspond to Euler angles
$\alpha = \phi_i$, $\beta = \theta_i$, and $\gamma = \varrho_i$ in the
notation of Scheme A of $\S$1.4.1 of Varshalovich, Moskalev, \&
Khersonskii (1988, hereafter VMK). From eq. (1) of their $\S$5.5.1 we
have
\begin{equation}
   Y_{\ell^\prime m^\prime}(\bfgamma^\prime) = 
   \sum^{\ell^\prime}_{m^{\prime\prime} = -\ell^\prime}
   Y_{\ell^\prime m^{\prime\prime}}(\bfgamma)\,
   D^{\ell^\prime}_{m^{\prime\prime}m^\prime}(\phi_i,\, \theta_i,\, 
   \varrho_i)\,,
   \label{ylm_rot}
\end{equation}
where $D^{\ell^\prime}_{m^{\prime\prime}m^\prime}$ is a Wigner
$D$-function corresponding to the rotation of the beam.  From
eqs. (\ref{ylm_rot}) and (\ref{beamrot}) we have
\begin{equation}
   b_{\ell m}(\bfgamma_i) = \sum^{\ell}_{m^\prime = -\ell}
   b_{\ell m^\prime}({\bf x}_3^\prime)
   D^\ell_{m m^\prime}(\phi_i,\, \theta_i,\, \varrho_i)\,.
   \label{blmrot}
\end{equation}

We focus on the elementary window function, i.e., ignore the
modulation and consider the window function for two points
$\bfgamma_i$ and $\bfgamma_j$ (see eq. [\ref{genwine}]).  Using the
usual decomposition for the Legendre polynomial in terms of spherical
harmonics, the fact that $B$ is real, and eqs.~(\ref{ylm_rot}) and
(\ref{blmrot}), we find
\begin{eqnarray}
   W_\ell^{({\rm e})}(\bfgamma_i,\, \bfgamma_j) & = & {4\pi \over 2\ell
   + 1} \sum^\ell_{m^\prime = - \ell} \, \sum^\ell_{m^{\prime\prime} =
   - \ell} \left[b_{\ell m^\prime}({\bf x}_3^\prime)\right]^* b_{\ell
   m^{\prime\prime}}({\bf x}_3^\prime) \nonumber \\ & {} & \times
   \sum^\ell_{m = - \ell} \left[D^\ell_{m m^\prime}(\phi_i,\,
   \theta_i,\,\varrho_i )\right]^* D^\ell_{m
   m^{\prime\prime}}(\phi_j,\, \theta_j,\, \varrho_j) .
   \label{wigwine1}
\end{eqnarray}
When the beam function is circularly symmetric $b_{\ell m}({\bf x}^\prime_3) = 
\delta_{m 0} B_\ell\,\sqrt{2\ell +1}$. Using $D^\ell_{m0}(\theta,\phi,
\varrho) = Y_{\ell m}(\theta,\phi)$, it is straightforward to establish
that in this case eq.~(\ref{wigwine1}) reduces to the usual expression
given in eq.~(\ref{isowine}).

Using the addition theorem for Wigner $D$-functions (eqs. [2] of
$\S$4.4, [1] of $\S$4.3, and [5] and [6] of $\S$4.7.2 of VMK),
we reduce eq. (\ref{wigwine1}) to the simpler form,
\begin{equation}
   W_\ell^{({\rm e})}(\bfgamma_i,\, \bfgamma_j) = {4\pi \over 2\ell +
   1} \sum^\ell_{m^\prime = - \ell} \, \sum^\ell_{m^{\prime\prime} =
   - \ell} \left[b_{\ell m^\prime}({\bf x}_3^\prime)\right]^* b_{\ell
   m^{\prime\prime}}({\bf x}_3^\prime) D^\ell_{m^\prime
   m^{\prime\prime}}(\alpha -\varrho_i,\, \gamma,\, \beta+\varrho_j),
   \label{wigwine}
\end{equation}
where
\begin{eqnarray}
   {\rm cos} \gamma & = & {\rm cos}\theta_i\, {\rm cos}\theta_j +  
   {\rm sin}\theta_i\, {\rm sin}\theta_j\, {\rm cos} (\phi_i - \phi_j)
   = \bfgamma_i \cdot \bfgamma_j \nonumber \\
   {\rm cot} \alpha & = & - {\rm cos}\theta_i\, {\rm cot}  (\phi_i - \phi_j) +  
   {\rm sin}\theta_i\, {\rm cot}\theta_j\, {\rm csc} (\phi_i - \phi_j) 
\label{wigwine_aux} \\
   {\rm cot} \beta & = & - {\rm cos}\theta_j\, {\rm cot}  (\phi_i - \phi_j) +  
   {\rm cot}\theta_i\, {\rm sin}\theta_j\, {\rm csc} (\phi_i - \phi_j) .
   \nonumber    
\end{eqnarray}

For large values of $\ell$ it is computationally expensive to evaluate
the entire $m^\prime$ and $m^{\prime\prime}$ sum in eq.~(\ref{wigwine}).  
However, for a smooth, mildly non-circular beam function (defined precisely
below), restricting the summation to a few low values of $m^\prime$ and 
$m^{\prime\prime}$ results in a good approximation. A smooth (i.e., not
sharply peaked) beam function results in $b_{\ell 0}$ 
falling off with increasing $\ell$.  At large $\ell$ and small $\gamma$, 
$D^\ell_{m m^\prime} (\alpha,\, \gamma,\, \beta)\rightarrow e^{-i m\alpha
-i m^\prime\beta} J_{|m-m^\prime|} ([\ell+1/2]\gamma)$. Thus the
$D^\ell_{m^\prime m^{\prime \prime}}$ term in eq.~(\ref{wigwine}) strongly 
suppresses, as $(\ell\gamma)^{|m^\prime-m^{\prime\prime}|}$, the 
contribution from off-diagonal $m^\prime \neq m^{\prime\prime}$ terms. In 
addition, mild non-circularity requires that at each value of $\ell$ the 
ratio $|b_{\ell m}/b_{\ell 0}|$ decrease rapidly with increasing $|m|$.
Hence the products $b_{\ell m^\prime}^* b_{\ell m^{\prime\prime}}$, as
functions of $\ell$, are ordered in magnitude and fall off as one goes
to higher values of $|m^\prime| + |m^{\prime\prime}|$. For Python V where 
the deviation from circularity is small, retaining the first non-zero order 
term is sufficient for computing an accurate covariance function.\footnote{ 
  By a sufficiently accurate covariance function we mean that the maximum 
  likelihood analysis results are not significantly affected by higher 
  order corrections. This criterion of sufficient accuracy therefore depends 
  on the level of noise in the experiment, which is encoded in the noise
  covariance matrix of the experiment.}

We now derive explicit expressions for the first few leading order
terms in eq.~(\ref{wigwine}) for the specific case of an elliptical
beam function. For an elliptical beam function, symmetry dictates that
$b_{\ell m}({\bf x}_3^\prime) = 0$ for odd $m$ (see
appendix~\ref{analbeam}). In what follows beam rotations are set to
zero ($\varrho_i = 0$), but it is straightforward to restore them from
the complete expression given above.  The zeroth order term contains
$D^\ell_{00}$, the first order term has four contributors
($D^\ell_{02}$, $D^\ell_{0,-2}$, $D^\ell_{20}$, and $D^\ell_{-2,0}$),
and the second order term has eight contributors ($D^\ell_{22}$,
$D^\ell_{2,-2}$, $D^\ell_{-2,2}$, $D^\ell_{-2,-2}$, $D^\ell_{04}$,
$D^\ell_{0,-4}$, $D^\ell_{40}$, and $D^\ell_{-4,0}$).

After a significant amount of algebraic manipulations (that use
equations from $\S\S$4.3, 4.4, 4.8, and 4.17 of VMK, familiar properties
of Legendre polynomials and modified Bessel functions, and the reality
condition on the beam, eq.~[\ref{beamreal}]), we find a series expansion
of the elementary window function,
\begin{eqnarray}
   W_\ell^{({\rm e})}(\bfgamma_i,\, \bfgamma_j) = {4\pi \over 2\ell + 1} 
   & \bigg[ & \left[b_{\ell 0} ({\bf x}_3^\prime)\right]^2 
                d^\ell_{00}(\gamma) + 2 b_{\ell 0} ({\bf x}_3^\prime) 
                b_{\ell 2} ({\bf x}_3^\prime)
                \left\{ {\rm cos}(2\alpha) + {\rm cos}(2\beta) \right\}
                d^\ell_{02}(\gamma) \nonumber \\
   & {} & + 2 \left[b_{\ell 2} ({\bf x}_3^\prime)\right]^2
              \left[ {\rm cos}(2\alpha + 2\beta) d^\ell_{22}(\gamma)
              + (-1)^{-\ell} {\rm cos}(2\alpha - 2\beta) 
              d^\ell_{22}(\pi - \gamma)\right] \nonumber \\
   & {} & + 2 b_{\ell 0} ({\bf x}_3^\prime) b_{\ell 4} ({\bf x}_3^\prime)
              \left\{ {\rm cos}(4\alpha) + {\rm cos}(4\beta) \right\}
              d^\ell_{04}(\gamma) + \cdots \bigg].
   \label{wigwin40}
\end{eqnarray}
Here the angles $\alpha$, $\beta$, and $\gamma$ are defined in
eqs.~(\ref{wigwine_aux}), and the $d^\ell_{mm^\prime}$'s are the usual
Wigner $d$-functions of angular momentum theory (e.g., $\S$4.3 of VMK)
related to the Wigner $D$-functions through $ D^\ell_{mm^\prime}
(\alpha,\gamma,\beta) = e^{-i m \alpha} \,d^\ell_{mm^\prime}(\gamma)
\, e^{-i m^\prime \beta}$. More precisely,
\begin{eqnarray}
   d^\ell_{00}(\gamma) & = & P_\ell({\rm cos}\gamma) , \nonumber \\
   d^\ell_{02}(\gamma) & = & - \sqrt{\ell(\ell + 1) \over (\ell - 1)
             (\ell + 2)} P_\ell({\rm cos}\gamma) + {2 {\rm cos}\gamma \over 
             \sqrt{(\ell - 1) \ell (\ell + 1) (\ell + 2)}} 
             P^\prime_\ell({\rm cos}\gamma) , \nonumber \\
   d^\ell_{22}(\gamma) & = & {1 \over (\ell - 1) (\ell + 2)} 
             \left[ -4 \left({2 - {\rm cos}\gamma \over 1 + {\rm cos}\gamma}
             \right) +  \ell(\ell + 1) \right] P_\ell({\rm cos}\gamma) 
             \nonumber \\
                      & {} & + {4(1 - {\rm cos}\gamma) \over (\ell - 1)
             \ell (\ell + 1) (\ell + 2)} \left[ - \left({2 - {\rm cos}\gamma 
             \over 1 + {\rm cos}\gamma}\right) +  \ell(\ell + 1) \right]
             P^\prime_\ell({\rm cos}\gamma) , 
\label{dlm_explicit}\\
   d^\ell_{22}(\pi - \gamma) & = & {(-1)^\ell \over (\ell - 1) (\ell + 2)} 
             \left[ -4 \left({2 + {\rm cos}\gamma \over 1 - {\rm cos}\gamma}
             \right) +  \ell(\ell + 1) \right] P_\ell({\rm cos}\gamma) 
             \nonumber \\
                           & {} & + {4(-1)^{\ell + 1} (1 + {\rm cos}\gamma) 
             \over (\ell - 1) \ell (\ell + 1) (\ell + 2)} \left[ - \left({2 +   
             {\rm cos}\gamma \over 1 - {\rm cos}\gamma}\right) +  
             \ell(\ell + 1) \right]  P^\prime_\ell({\rm cos}\gamma) , 
             \nonumber \\
   d^\ell_{04}(\gamma) & = & - \ell(\ell + 1) \sqrt{(\ell - 4)! \over 
             (\ell + 4)!} \left[{6 \left(1 + 3 {\rm cos}^2\gamma\right) 
             \over {\rm sin}^2\gamma} - \ell(\ell + 1) \right] 
             P_\ell({\rm cos}\gamma) \nonumber \\
                      & {} & + 8 \sqrt{(\ell - 4)! \over (\ell + 4)!}\,
             {\rm cos}\gamma \left[{3 \left(1 + {\rm cos}^2\gamma\right) 
             \over {\rm sin}^2\gamma} - \ell(\ell + 1) \right]
             P^\prime_\ell({\rm cos}\gamma) \,, \nonumber
\end{eqnarray}
where $P^\prime_\ell\equiv dP_\ell(x)/dx$. Recursion relations in the
indices $\ell$, $m$ and $m^\prime$ (see VMK) can be used to compute
$d^\ell_{mm^\prime}$ for larger values of $m$ and $m^\prime$.

Evaluating the first few terms of the Wigner method expansion in 
eq. (\ref{wigwin40}) involves computing $P_\ell$ and $P^\prime_\ell$.
The first derivative $P^\prime_\ell$ can be readily computed in terms
of $P_\ell$ and $P_{\ell-1}$ during the generation of $P_\ell$ using
the upward recursion relation. Hence the computational cost of evaluating 
$W_\ell^{({\rm e})}$ for a mildly non-circular beam function (using eq. 
[\ref{wigwin40}]) is only a factor of a few larger than that for a 
circular beam function.

Figure 2 shows, as a function of $\ell\sigma_{\rm RMS}$, the six leading 
order $|b_{\ell m}({\bf x}_3^\prime)\,b_{\ell m^\prime}({\bf x}_3^\prime)|$  
coefficients (with $m,m^\prime = 0,2,4$) of the expansion of eq. (\ref{wigwine})
(see eq. [\ref{wigwin40}]) for the Python V experiment. These are also the 
leading order contributors to the zero-lag elementary window function. Note
that the curves do not cross at any $\ell$, i.e., the ordering of the 
coefficients is maintained for all $\ell$, and at large $\ell$ (past the peak) 
higher order terms fall off more rapidly with $\ell$. Hence this perturbation
expansion is an efficient scheme for computing non-circularity corrections.
The non-circularity corrections peak at angular scales smaller than the RMS 
beamwidth $\sigma_{\rm RMS}$. Thus non-circularity corrections are not that 
important for an elementary window function (a single beam experiment) but 
can have a significant effect on the complete window function for a
modulated experiment if the modulation scheme results in sensitivity to the
$\ell\sigma_{\rm RMS} > 1$ regime.

Figure 3 shows contour plots of the isotropic term $d^\ell_{00}(\gamma)$ 
and the leading order correction term $\left\{ {\rm cos}(2\alpha) + 
{\rm cos}(2\beta) \right\} d^\ell_{02}(\gamma)$ in the perturbation 
expansion of the Wigner method elementary window function (see 
eq.~[\ref{wigwin40}]) for the Python V experiment. These are plotted for 
multipole $\ell=100$ chosen so that $\ell \sigma_{\rm RMS} \sim 1$, which is 
where the non-circularity correction starts becoming significant.
Close to the center of the plots this non-circularity correction is
larger along the major and minor axes of the elliptical beam. The
correction term falls off with increasing pixel separation, suggesting
that the circular beam function approximation is good for sufficiently
large separations. Also, for a modestly non-circular beam function
this implies that higher order terms in the Wigner method perturbation
expansion need be retained only for close pixel pairs and one can
hence truncate the summation in eq. (\ref{wigwine}) at lower orders
for progressively more widely separated pixel pairs.

\section{Constant Elevation Scan Window Function}
\label{wf_eqdec}

For an experiment, such as Python V, that scans at constant elevation\footnote{ 
  Constant elevation here refers most generally to any set of 
  parallel circles on the sky.}, 
it is possible to derive another expression for the window function which
does not require use of the approximation of the previous section (the
truncation of the $m^\prime$ and $m^{\prime\prime}$ series).

We follow the initial development of the previous section and transform to 
a new coordinate system by rotating around the ${\bf x}_3$ axis by $\phi_i$, 
where $\bfgamma_i$ $=(\theta_i,\, \phi_i)$ is the pointing direction. The 
Euler angles of this rotation are $\alpha = \phi_i$, $\beta = 0$, and 
$\gamma = \varrho_i$. As described in the previous section, $\varrho_i$ 
represents a relative rotation of the beam around its axis. For an experiment 
located at one of the Poles $\varrho_i=0$. In general, we have
\begin{equation}
   b_{\ell m}(\bfgamma_i) = \sum^{\ell}_{m^\prime = -\ell} b_{\ell
   m^\prime}({\bf x}_3^\prime) D^\ell_{m m^\prime}(\phi_i,\, 0,\,
   \varrho_i) .
   \label{blmrot_cdec}
\end{equation}
From $\S$4.16 of VMK we find $D^\ell_{m m^\prime}(\phi_i,\, 0,\, \varrho_i)
= e^{-im(\phi_i + \varrho_i)}\,\delta_{mm^\prime}$. Equation
(\ref{blmrot_cdec}) then implies $b_{\ell m}(\theta_i,\, \phi_i) =
e^{-im(\phi_i+ \varrho_i)} b_{\ell m}(\theta_i,\, 0)$ and so
eq. (\ref{wigwine1}) reduces to
\begin{equation}
   W_\ell^{({\rm e})}(\bfgamma_i,\, \bfgamma_j) = {4\pi \over 2\ell +
   1} \sum^\ell_{m = - \ell} e^{-im\left[(\phi_i -
   \phi_j)+(\varrho_i-\varrho_j)\right]}\,\, b^*_{\ell m}(\theta_i,\, 0)
   \,b_{\ell m}(\theta_j,\, 0) .
   \label{cdecwine1}
\end{equation}
Window functions between pixels lying on a few constant elevation lines
can be rapidly computed by using eq. (36) and pre-computed $b_{\ell m}
(\theta_i,\, 0)$'s.

For an experiment like Python V whose beam function has the symmetry
$B(\theta_i ,\, 0 ;\, \theta ,\, \phi) = B(\theta_i ,\, 0 ;\, \theta ,\, 
- \phi )$, it may be shown that eqs. (12) and (26) and the fact that the 
beam function $B$ is real implies that $b_{\ell m}(\theta_i,\, 0)$ is real.
In this case eq. (36) may be re-expressed as
\begin{equation}
   W_\ell^{({\rm e})}(\bfgamma_i,\, \bfgamma_j) = {4\pi \over 2\ell +
   1} \sum^\ell_{m = - \ell} {\rm cos} \left[m \{(\phi_i -
   \phi_j)+(\varrho_i-\varrho_j)\}\right]\,\, b_{\ell m}(\theta_i,\, 0)
   \,\,b_{\ell m}(\theta_j,\, 0) .
   \label{cdecwine2}
\end{equation}
Here $b_{\ell m}(\theta,\, 0)$ is defined in the usual way through
eq.~(\ref{blm}). For the Python V experiment, which chops at constant elevation 
and where the pixels lie on a relatively small number of elevations\footnote{
  The 690 sky pixels of Python V lie on $11$ distinct elevations.}, it is 
possible to pre-compute and store the $b_{\ell m}(\theta,\, 0)$ at all 
elevations $\theta$. Given these pre-computed $b_{\ell m}$'s, the
elementary window function can be computed very rapidly by using
eq.~(\ref{cdecwine2}).

\section{Comparison of Approximate and Exact Python V Covariance Matrices}
\label{wincomp}

Python V is a CMB anisotropy experiment which performs wide-angle scans with 
a non-circular beam. In this section we compare Python V window functions 
and theoretical covariance matrices computed in the flat-sky approximation 
and in the Wigner method perturbative approximation as well as in the exact 
method.

Python V observations were made at 37--45 GHz.  Two regions of the sky were 
observed: the main field, a rectangle $7.\!\!^\circ 5$ in declination 
($\delta=-52^\circ$ to $-45.\!\!^\circ 4$) by $67.\!\!^\circ 7$ in azimuth 
(centered on $\alpha = 23.\!\!^{\rm h} 18$); and another rectangular patch
$3^\circ$ in declination ($\delta=-63^\circ$ to $-60^\circ$) by $30^\circ$
in azimuth (centered on $\alpha =3.\!\!^{\rm h} 0)$. For detailed descriptions
of the experiment and data see Coble et al. (1999) and Coble (1999).

The Python V beam function is well described by an elliptical Gaussian
with FWHM beam\-widths of $1.\!\!^\circ 02^{+0.\!\!^\circ 03}_{-0.\!\!^\circ 
01}$ in elevation and $0.\!\!^\circ 91^{+0.\!\!^\circ 03}_{-0.\!\!^\circ 01}$ 
in azimuth (one standard deviation uncertainties). The Python V beam function 
is a compact elliptical Gaussian, eq.~(\ref{angau_beam}), with $\sigma_1= 
0.0076$ and $\sigma_2=0.0067$ as the nominal beamwidths in radians. Python V
uses a constant-elevation smooth scan sampling strategy around every pixel 
on the sky, with a chopper throw (end to end) of $\Phi_c = 17.\!\!^\circ 06$. 
Each chopper cycle consists of $128$ time samples suitably modulated in
time to correspond to the spatial modulations described below. To compute the 
window function we only need to know the final spatial modulation strategy
adopted. See Coble (1999) for a more detailed discussion of these procedures.
The constant elevation scans are discretely resampled in space (with nine-fold 
oversampling) at $N_c=567$ equi-spaced points $(\theta_i,\phi^p_i)$ labelled by 
integers $p = 1,2,\ldots,N_c$ along
the chopper cycle around the pixel $\bfgamma_i = (\theta_i, \phi_i)$.
The azimuth $\phi^p_i = \phi_i + (p - 1) \Delta\phi +(\Phi_o
-\Phi_c/2)$, where $\Delta\phi = \Phi_c/(N_c-1) $ is the spacing
between points and $\Phi_o = 0.\!\!^\circ 58$ accounts for the offset
between the azimuth of the pixel, $\phi_i$, and the center of scan.
The scans are modulated using the first eight cosine harmonics of the
chopper cycle (hereafter modulations 1 to 8).  All modulations, other
than the first, are apodized by a Hann window to reduce ringing in
multipole space and down weight data taken during chopper turnaround. 
For the modulated Python V scans the weight functions (see eq. 
[\ref{wtcombtemp}]) are\footnote{ The following expressions
  correct a typographical error in the corresponding expressions in
  Coble et al. (1999) and Coble (1999).}
\begin{equation}
   w^{(m)}_{i p} = {2 M^{(m)}_p \over \sum_{p=1}^{N_c} |M^{(m)}_p|}
   \label{pyv_modwt}
\end{equation}
where 
\begin{eqnarray}
   M^{(1)}_p & = & \cos(2\pi Z_p) \nonumber \\
   M^{(m)}_p & = & \frac{(-1)^{m+1}}{2}\,\cos(2\pi m Z_p)\,
   \left[1-\cos(2\pi Z_p)\right], ~~~~(m > 1) 
   \label{pyv_mod}
\end{eqnarray}
with $Z_p = (p-1)/(N_c -1)$.  These weights are used to obtain the
Python V complete window function $W_\ell^{(n,m)}(\bfgamma_i,\, \bfgamma_j)$
(between modulation $m$ of the scan around sky pixel $\bfgamma_i$ and
modulation $n$ of the scan around sky pixel $\bfgamma_j$) in terms of
the elementary window functions (see eq.~[\ref{win_wine}]).

The $M^{(m)}_p$ are identical for all pixels hence $w^{(m)}_{i p}$ is 
independent of pixel index $i$. With identical constant-elevation chops 
around every pixel, eq.~(\ref{win_wine}) can be re-expressed in a 
computationally more
efficient form. In this case the elementary window function
$W_\ell^{({\rm e})}$ depends only on $\phi^p_i - \phi^q_j$ 
(at fixed $\theta_i$ and $\theta_j$). This reduces the number of
separations at which $W_\ell^{({\rm e})}$ is needed and thus speeds up
the computation. In this case we may re-express the window function as
\begin{eqnarray}
   W_\ell^{(n,m)}(\bfgamma_i,\, \bfgamma_j) = & {} &
   \sum^{N_c-1}_{q_1=1} \sum^{N_c-q_1}_{q_2=1} \Bigg[ w^{(n)}_{i (q_2+q_1)}
   w^{(m)}_{j q_2} W_\ell^{({\rm e})} \left(\theta_i,\, \theta_j,\,
   \phi_i - \phi_j + q_1\Delta\phi\right) \nonumber \\ & {} & \ \ \ \
   \ \ \ + w^{(n)}_{i q_2} w^{(m)}_{j (q_2+q_1)} W_\ell^{({\rm e})}
   \left(\theta_i,\, \theta_j,\, \phi_i - \phi_j -
   q_1\Delta\phi\right) \Bigg] \nonumber \\ & {} & + 
   \sum^{N_c-1}_{q_2=1} w^{(n)}_{i q_2} w^{(m)}_{j q_2} W_\ell^{({\rm e})}
   \left(\theta_i,\, \theta_j,\, \phi_i - \phi_j\right)\,.
\label{win_wine_eff}
\end{eqnarray}
Here the three arguments of the elementary window functions (the three
$W_\ell^{({\rm e})}$ between two directions) are the colatitudes of the two 
directions and the difference in their azimuth angles. Pre-computing the 
second summations over the weight products, $\varpi^{(m,n)}_p=
\sum^{N_c-p}_{q=1} w^{(n)}_{i (q+p)} w^{(m)}_{j q}$, for the set of all the 
modulation pairs speeds up the evaluation of the right hand side of
eq.~(\ref{win_wine_eff}).

Figure 4 shows the eight equal-modulation exact Python V zero-lag
complete window functions at the two extreme values of the
declination, $\delta = - 63^\circ$ and $\delta = -45.\!\!^\circ 4$.
Also plotted for comparison is the zero-lag elementary window function
at $\delta=-63^\circ$. The effect of beam function non-circularity is
more pronounced for the modulated (complete) window functions since
these have peak sensitivity at multipoles well beyond the inverse
beamwidth, which is where the non-circularity corrections start to
become important (see Fig. 2).

While it is of interest to estimate the accuracy of window functions
computed in various approximations, the accuracy of computed covariance
matrix elements is of much greater relevance since these directly determine 
the accuracy of cosmological results extracted from CMB anisotropy data.
This  significantly extends the range of usefulness of approximate
window functions. First, the sum over $\ell$ in the definition of the 
theoretical model covariance matrix, eq. (\ref{ctij_wl}), averages
over and hence reduces the significance of deviations between the
approximate and exact window functions. For example, an approximate
window function that has large deviations from the exact window function 
only in regimes where they oscillate in $\ell$ may still result in an 
accurate covariance matrix. Second, errors in the window function are 
unimportant when the corresponding covariance matrix elements are small 
(subdominant). For example, while the flat-sky approximation window 
function is very inaccurate for a widely separated pixel pair, the covariance 
matrix element for such a pixel pair is subdominant and thus cannot 
significantly influence the results from a maximum likelihood analysis of the 
data. Another important consideration is the level of noise in the experiment. 
Since the inverse of the sum of the theoretical model and noise covariance 
matrices, $(C_T + C_N)^{-1}$, determines the results of the maximum likelihood 
analysis, the model covariance matrix needs to be computed to higher accuracy 
for an experiment with lower noise.

Equation~(\ref{ctij_wl}) defines the model covariance matrix 
element $C_{ij}^{(m, n)}$ in terms of the window function and the model CMB 
anisotropy power spectrum. To compare covariance matrices computed using 
different approximations therefore requires choice of a model $C_\ell$. We use 
the flat bandpower spectrum, $C_\ell \propto 1/\ell(\ell+1)$, for this purpose
in what follows. With this choice of power spectrum, the quantity
\begin{equation}
   F_\ell^{(n,m)}(\bfgamma_i, \bfgamma_j) =
   \sum_{\ell^\prime =2}^\ell {2\ell^\prime+1 \over \ell^\prime(\ell^\prime+1)} 
   W_{\ell^\prime}^{(n,m)}(\bfgamma_i,\, \bfgamma_j)\,,
   \label{diffcorr}
\end{equation}
is a measure of the cumulative build up of the corresponding covariance
matrix element as one progresses with the sum over multipole $\ell$ in 
eq.~(\ref{ctij_wl}). Up to a multiplicative normalization constant, 
$F_\ell^{(n,m)}(\bfgamma_i, \bfgamma_j)$ converges to the covariance 
matrix $C_{ij}^{(n,m)}$ as $\ell \to \infty$. Unlike $W_\ell$, $F_\ell$ 
measures $\ell$-space information that may be directly used to estimate 
the accuracy of the computed covariance matrix. It also allows one to 
determine the value of $\ell$ to which one must compute to achieve 
a desired accuracy.

Figure 5 compares Python V flat bandpower covariance matrix elements
computed using the different approximations developed above. The upper
panels, $a)$ and $b)$, show comparisons of zero-lag covariance matrix
elements.\footnote{
  Modulated zero-lag complete window functions receive contributions 
  from non-zero-lag elementary window functions between pixels separated 
  by as much as the chopper throw $\Phi_c = 17.\!\!^\circ 06$.}
At zero-lag the flat-sky approximation is more accurate than
the zeroth order (circular beam) and first order (wig20) Wigner method
perturbation expansion approximations. The Wigner method gets progressively 
more efficient (i.e., one needs
to retain fewer terms in the perturbation expansion series to achieve
the desired accuracy) for non-zero-lag covariance matrix elements
between increasingly separated pixels.  For covariance matrix
elements between very widely separated pixels even the circular beam
(zeroth order) approximation suffices. The Wigner method is more
accurate for lower modulations which probe lower values of $\ell$
where beam non-circularity corrections are smaller (see Fig.  5).  The
flat-sky approximation is accurate at small separations (e.g., at
zero-lag).  For Python V the flat-sky approximation is more accurate for
pixels separated in azimuth than for pixels separated in declination.
The flat-sky approximation works better for higher modulations which 
probe larger values of $\ell$. The $F_\ell$ curves for the circular beam 
approximation (wig00) in the lower left panel $c)$ also highlights the 
possible pitfall of not computing to large enough $\ell$. Here $F_{\ell ij}$ 
does not converge to $C_{ij}$ until $\ell\gsim 500 \sim 4\sigma_{\rm RMS}^{-1}$.

\section{Conclusion}
\label{concl}

We develop computationally rapid methods to compute the window function 
for a long-scan arbitrary beam shape CMB anisotropy experiment. We use 
these methods to compute the window function for the elliptical Gaussian
beam Python V experiment.

It proves convenient to separate effects due to the modulation scheme adopted 
and the shape of the beam function by expressing the complete window function 
as a weighted sum of single-beam elementary window functions, 
eq.~(\ref{win_wine}).

Using eq. (11) to obtain exact elementary window function for a non-circular 
beam experiment requires accurate computation of the spherical harmonic 
transform of the beam function at each pointing direction. For an 
experiment with a large number of pointing directions this is computationally
prohibitive. For instance, Python V has 690 pixels and the scan around
each is resampled at 567 points which results in $\approx 0.4$ million
pointing directions. Fortunately, the 690 pixels lie on only 11 distinct 
elevations and the scans are performed at constant elevation. Hence
pre-computing and storing the spherical harmonic beam function transforms 
at these 11 declinations allows for rapid computation of the exact Python V 
window function.

In the absence of such a simplification due to a symmetry, the Wigner
method perturbation expansion scheme allows an accurate computation of
the window function with computational effort within a factor of a few
of the corresponding computation for the case of a circular beam
function. This factor depends on the order to which the perturbative
expansion (about the circular beam approximation) needs to be developed to 
achieve the desired accuracy. In this implementation, the Wigner method 
requires pre-computation and storage of one spherical harmonic transform of
the beam pointing at a Pole, $b_{\ell m}(\bfgamma_{\rm P})$. If the beam 
is mildly non-circular then for all $\ell$ $|b_{\ell m}(\bfgamma_{\rm P})|/
|b_{\ell 0}(\bfgamma_{\rm P})|$ falls off rapidly with increasing $|m|$, 
allowing for a fast and accurate transform and simpler storage. In the 
Appendix we record a semi-analytic expression for the beam function 
transform for a compact elliptical Gaussian beam. In mildly non-circular 
cases, such as Python V, $|b_{\ell m}(\bfgamma_{\rm P})|/
|b_{\ell 0}(\bfgamma_{\rm P})|\to 0$ rapidly with increasing $|m|$ and 
the first order Wigner method is sufficiently accurate. We also develop a 
flat-sky approximation for window function computation and illustrate this
method by computing the Python V window function.\footnote{
  Our flat-sky approximation differs from that implemented in Coble 
  et al.  (1999) and Coble (1999).} 
We find that the flat-sky approximation works well at zero-lag and for pixels 
at small constant-elevation separation. At larger separations the Wigner 
method is more accurate than the flat-sky approximation.

The methods developed in this paper are easily extended to other cases not 
explicitly considered here (such as a non-circular non-Gaussian beam, a 
beam that rotates on the sky, etc.). Elsewhere we summarize an analysis of 
the Python V data that makes use of these methods.

\bigskip

We acknowledge very valuable discussions with K. Coble, advice from 
S. Dodelson, M. Dragovan, K. Ganga, L. Knox, and the rest of the Python
collaboration, and support from NSF CAREER grant AST-9875031.

\appendix

\section{Elliptical Gaussian Beam Function: Normalization \& Transforms}
\label{analbeam}

Python V is an example of an experiment with a compact elliptical Gaussian 
beam function (Coble et al. 1999). In such a case it is possible to
obtain accurate and useful semi-analytic expressions for the Fourier 
transform $B({\bf k})$ and the spherical harmonic transform $b_{\ell 
m}(\bfgamma_{\rm P})$ of the beam function.

An elliptical Gaussian beam function that is compact enough can be expressed, 
in a locally flat-sky coordinate system (around the beam pointing direction),
as
\begin{equation}
   B({\bf x}) = {1 \over 2\pi\sigma_1\sigma_2} {\rm exp} \left[ -
   {x_1{}^2 \over 2\sigma_1{}^2} - {x_2{}^2 \over 2\sigma_2{}^2} \right]\,.
   \label{angau_beam}
\end{equation}
Here ${\bf x} = (x_1,\, x_2)$ are locally flat-sky cartesian 
coordinates and $\sigma_1$ and $\sigma_2$ are the beamwidths in the
$\bf x_1$ and $\bf x_2$ directions.\footnote{ 
   The beam function is also a function of the pointing direction 
   $\bf x_0$, $B({\bf x_0},\, {\bf x})$, taken to be at the origin here.}
It is straightforward to establish that this is normalized so that
\begin{equation}
   \int^\infty_{-\infty} dx_1 \int^\infty_{-\infty} dx_2 \, B({\bf x}) = 1\,.
\end{equation}

We shall have need for the expression for the beam in local spherical
polar coordinates ${\bfgamma},\, =(\theta ,\, \phi)$, around the
pointing direction which, without loss of generality, can be assumed
to be the North Pole $\bfgamma_{\rm P}$. To fix the orientation of the
beam on the sky we choose $x_1$ to lie along $\phi = 0$. Using the
local mapping, $x_1 = \theta {\rm cos} \phi$ and $x_2 = \theta {\rm sin}
\phi$, we can write
\begin{equation}
   B(\bfgamma_{\rm P},\bfgamma) = {1 \over 2\pi \sigma_1 \sigma_2}
   {\rm exp} \left[ - {\theta^2 \over 2 \sigma^2 (\phi)} \right] ,
   \label{ellipbeam_polar}
\end{equation}
where the ``beamwidth'' is a function of the polar angle
\begin{equation}
   \sigma^2 (\phi) = {\sigma_1{}^2 \over 1 + \epsilon^2 {\rm sin}^2 \phi}\,,
\end{equation}
and a  non-circularity parameter
\begin{equation}
   \epsilon = {\sigma_1{}^2 - \sigma_2{}^2 \over \sigma_2{}^2}\,.
\end{equation}
With the usual accurate approximation, allowed by the rapid fall off
of the Gaussian in $\theta$ in eq.~(\ref{ellipbeam_polar}), it is
straightforward to establish that the normalization condition
\begin{equation}
   \int^{2\pi}_0 d\phi \int^\pi_0 d\theta\, {\rm sin}\theta\, 
   B(\bfgamma_{\rm P}, \bfgamma) = 1
\end{equation}
is satisfied for $\sigma(\phi) \ll 1$.

In the flat-sky approximation the elementary window function defined in 
eq. (\ref{flatwine}) depends on the Fourier transform of the beam function
$B({\bf k})$. For the elliptical Gaussian beam function of 
eq.~(\ref{angau_beam}) we find
\begin{equation}
   B({\bf k}) = {\rm exp}\left[ - {k_1{}^2 \sigma_1{}^2 \over 2}  
   - {k_2{}^2 \sigma_2{}^2 \over 2} \right] \,.
   \label{pyv_beam_ft}
\end{equation}
One great advantage of the flat-sky approximation is that Fast Fourier 
Transform techniques may be used to rapidly compute $B({\bf k})$ for  
any beam function.

To use eq.~(\ref{wigwine}) to compute the curved-sky window function
we need to compute the spherical harmonic transform $b_{\ell m}
(\bfgamma_{\rm P})$ of the beam function pointed at the North Pole
$\bfgamma_{\rm P}$. An elliptical Gaussian beam function results in a
semi-analytic expression that requires numerical evaluation of only a
single integral. A more complex beam function could require a complete
numerical analysis.  

We first note that it is straightforward to show that
\begin{equation}
   \left[b_{\ell m}(\bfgamma)\right]^* = (-1)^m b_{\ell , -m}(\bfgamma)\,.
\label{beamreal}
\end{equation}
From eqs.~(\ref{ellipbeam_polar}) and (\ref{blm}) we find
\begin{equation}
   b_{\ell m}(\bfgamma_{\rm P}) 
   = {(-1)^{-m} \over 2\pi\sigma_1 \sigma_2} \sqrt{ {2\ell + 1 \over 4\pi} 
   {(\ell + m)! \over (\ell - m)!} }
   \int^{2\pi}_0 d\phi\, e^{-im\phi}
   \int^\pi_0 d\theta\, {\rm sin}\theta\, 
   e^{-\theta^2/[2 \sigma^2(\phi)]} P^{-m}_\ell ({\rm cos}\theta)\,,
\end{equation}
where we have used the usual expression for $Y_{\ell m}^*$ in terms of
the associated Legendre function $P^{-m}_\ell$. For a compact beam, and
for large $\ell$, the $\theta$ integral may be performed using the
usual small-$\theta$ approximation,
\begin{equation}
  {\rm lim}_{\ell \rightarrow \infty} \ell^m P^{-m}_\ell ({\rm cos}\theta)
  = J_m \left([\ell + 0.5] \theta\right) ,
\end{equation}
e.g., eq. (8.722.2) of Gradshteyn \& Ryzhik (1994). Using eq. (6.631.7) of 
Gradshteyn \& Ryzhik (1994) we find
\begin{eqnarray}
   b_{\ell m}(\bfgamma_{\rm P}) 
   & = & {(-\ell)^{-m} \over \pi (\ell + 0.5)^{3/2}} \sqrt{ 
         {(\ell + m)! \over (\ell - m)!} } {1 \over \sigma_1 \sigma_2}   
\label{blm_anal} \\
   & {} & \times \int^{2\pi}_0 d\phi\, e^{-im\phi} f^3(\phi) e^{-f^2(\phi)}
          \left[I_{(m-1)/2}\left(f^2(\phi)\right) - 
          I_{(m+1)/2}\left(f^2(\phi)\right)\right] , \nonumber
\end{eqnarray}
where $I_\nu$ is the modified Bessel function and 
\begin{equation}
   f(\phi) = {(\ell + 0.5) \sigma(\phi) \over 2} .
\end{equation}
Equation~(\ref{blm_anal}) is valid for $m \geq 0$; for $m < 0$ we use
this and eq.~(\ref{beamreal}) for the $b_{\ell m}$'s. It is
straightforward to show that for a circular beam eq.~(\ref{blm_anal})
reduces to the well known expression.

Using the reality condition on the beam, eq.~(\ref{beamreal}), one may
show that the $\int^{2\pi}_0 d\phi$ integral in eq.~(\ref{blm_anal})
may be replaced by a $\left[ 1 + (-1)^{-m}\right] \int^{\pi}_0 d\phi$
integral.  Clearly, the $b_{\ell m}({\bf x}_3^\prime)$'s vanish for
odd $m$.\footnote{This is true provided the integral does not
  diverge. In fact it is straightforward to establish that it is
  integrable. We use $e^{im\phi} = {\rm cos} (m\phi) + i {\rm sin}
  (m\phi)$ and consider the real and imaginary parts separately. Both
  of these are continuous functions of $\phi$ over $0 < \phi < \pi$
  and are thus Riemann integrable over this interval (see pp.  42 and
  63 of Whittaker \& Watson 1969).}  
Since $f(\phi)$ is a function of ${\rm sin}^2(\phi)$, it is straightforward 
though tedious to show that the imaginary part of $e^{-im\phi}$ in 
eq.~(\ref{blm_anal}) leads to an expression that vanishes. Thus we have
\begin{eqnarray}
   b_{\ell m}(\bfgamma_{\rm P}) 
   & = & {\left[ 1 + (-1)^{-m}\right] (\ell)^{-m} \over \pi(\ell + 0.5)^{3/2}}
         \sqrt{{(\ell + m)! \over (\ell - m)!} } {1 \over \sigma_1 \sigma_2} \\
   & {} & \times \int^{\pi}_0 d\phi\, {\rm cos}(m\phi) f^3(\phi) e^{-f^2(\phi)}
          \left[I_{(m-1)/2}\left(f^2(\phi)\right) - 
          I_{(m+1)/2}\left(f^2(\phi)\right)\right] . \nonumber
\end{eqnarray}
For an elliptical Gaussian beam function, this approximate semi-analytic 
spherical harmonic transform agrees well with the exact fully numerical 
transform.

\clearpage

\centerline{\bf FIGURE CAPTIONS}

\figcaption[]{Contour plots of the zeroth order (eq. [22], left panel)
  and first order (eq. [23], right panel) terms in the non-circularity
  perturbation expansion of the flat-sky approximation elementary
  window function for an elliptical Gaussian beam function experiment.
  These are computed for the nominal FWHM beamwidths of the Python V
  experiment, $1.\!\!^\circ 02$ in elevation and $0.\!\!^\circ 91$ in
  azimuth. They are plotted as a function of dimensionless variables 
  $ k{\bf x}$, and the two panels are centered on the center 
  of the zeroth order circular beam function window function.  As expected,
  the flat-sky window function for the circular beam in the left panel
  is circularly symmetric. For a fixed value of $k$ the first order
  correction in the right panel must be multiplied by $ -
  k^2(\sigma_2^2 -\sigma_1^2)$ before being added to the zeroth order
  term. For Python V the higher order terms are small compared to the
  first order term and visually have roughly similar structure.}

\figcaption[]{Coefficients $|b_{\ell m}({\bf x}_3^\prime) \, 
   b_{\ell m^\prime}({\bf x}_3^\prime)|$ of the six lowest order terms
   in the perturbation expansion of the Wigner method elementary window 
   function (see eqs.~[\ref{wigwin40}] and [\ref{wigwine}]). These are
   computed for the elliptical Gaussian Python V beam function and are 
   plotted as a function of $\ell\sigma_{\rm RMS}$. For the Python V 
   beam function the ellipticity parameter $\epsilon = 0.26$.
   Non-circularity corrections are important for $\ell\sigma_{\rm RMS} > 1$. 
   Note that the peak shifts to higher values of $\ell$ for higher order terms,
   relative to the peak position for lower order coefficients. The shape
   of the curves are independent of $\sigma_{\rm RMS}$ but depend 
   sensitively on the beam function ellipticity.}

 \figcaption[]{Contour plots in the azimuth-declination plane (with
   azimuth along the horizontal axis) of terms in the Wigner method
   perturbation expansion of the elementary window function for the
   Python V experiment.  These are plotted for multipole $\ell = 100$ such
   that $\ell\sigma_{\rm RMS} \sim 1$.  The left panel shows the zeroth
   order isotropic term $d^\ell_{00}(\gamma)$ and the right panel
   shows the first order correction term $\left\{ {\rm cos}(2\alpha) +
     {\rm cos}(2\beta) \right\} d^\ell_{02}(\gamma)$ (see eq. [33]).
   Here $\gamma$ is the angular separation between the central pixel
   and the pixel at the given azimuthal and declination sky
   coordinates.}

 \figcaption[]{Two sets of the eight equal-modulation exact Python V
   zero-lag complete window functions $W^{(m,m)}_\ell
   (\bfgamma,\bfgamma)$. The dark solid and light dotted curves
   correspond to window functions at the two extreme declinations
   $\delta=-63^\circ$ and $-45.\!\!^\circ 4$, respectively. Higher
   modulation window functions peak at progressively larger values of
   $\ell$ and with smaller amplitude. The dashed curve is the exact
   Python V zero-lag elementary window function at $\delta=-63^\circ$.
   Note that higher modulation complete window functions peak at
   $\ell$ a few times larger than the inverse beamwidth
   ($\sigma^{-1}_{\rm RMS} \approx 140$) which is where the
   non-circularity corrections start to become important (see Fig.~2).
   For the same difference in azimuth the angular separation between
   two equal declination points ($\Delta\phi\cos\delta$) is
   smaller at larger $|\delta|$. Consequently, for the same
   modulation, the window function at $\delta=-63^\circ$ peaks at a larger
   multipole than the window function at $\delta=-45.\!\!^\circ 4$.}

 \figcaption[]{Comparisons between Python V flat bandpower
   equal-modulation covariance matrix elements computed using
   different approximations.  Plotted are the relative difference
   between $F^{(m,m)}_\ell(\bfgamma_i, \bfgamma_j)$ (eq. [41])
   computed using an approximate window function and computed using
   the exact window function, and normalized by dividing by the flat
   bandpower $C^{(m,m)}_{ij}$. The approximations considered are the
   circular beam approximation, (``wig00'' in blue), the three
   successive leading improvements to this in the Wigner method
   perturbation expansion (``wig20'' in green, ``wig22'' in red, and
   ``wig40'' in black, see eq.  [33]; here each successive improvement
   includes all lower order terms), and the flat-sky approximation
   (``flat'' in cyan).  Two curves are shown for each approximation,
   those corresponding to modulation 2 (dashed) and 8 (solid). At large
   $\ell$ the modulation 8 curves converge to a lower accuracy than
   the corresponding modulation 2 curves because the non-circularity
   correction is more significant for higher modulations. Upper panels 
   show differences for zero-lag covariance matrix elements at the two 
   extreme declinations: $a)$ $\delta = -63^\circ$ and $b)$ $\delta = 
   -45.\!\!^\circ 4$. Lower
   panels show differences for non-zero-lag covariance matrix
   elements.  Panel $c)$ corresponds to pixels separated in azimuth by
   $20^\circ$ at declination $\delta = -63^\circ$ and panel $d)$
   corresponds to two neighboring, equal azimuth, pixels at
   declinations $\delta = -63^\circ$ and $-62^\circ$ (in panel $d$ the 
   solid red curve covers the solid black curve for $\ell \lsim 300$). 
   The flat-sky approximation fares well in all cases except for pixels 
   separated in declination, where it fails even for small-separation pixel
   pairs, see panel $d)$. Panel
   $d)$ also shows the enormous improvement over the circular beam
   function approximation (wig00) achieved by accounting for even just
   the first order Wigner method correction (wig20).  Panel $c)$
   highlights the need to compute to a large enough value of $\ell$ to
   achieve sufficient accuracy, e.g., truncating the circular beam
   function approximation (wig00) at intermediate $\ell$ leads to an
   inaccurate result.}

\clearpage

\begin{figure}[p]
  \resizebox{0.49\textwidth}{!}{\includegraphics{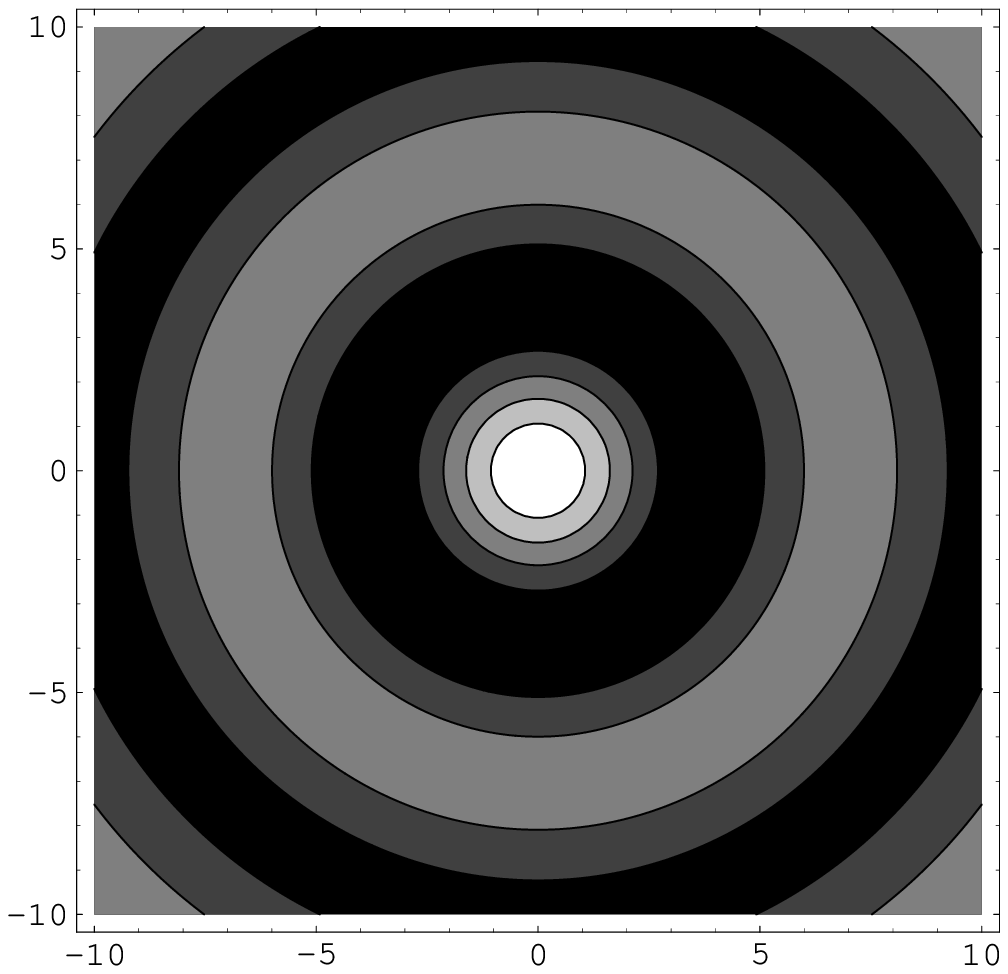}}
  \resizebox{0.49\textwidth}{!}{\includegraphics{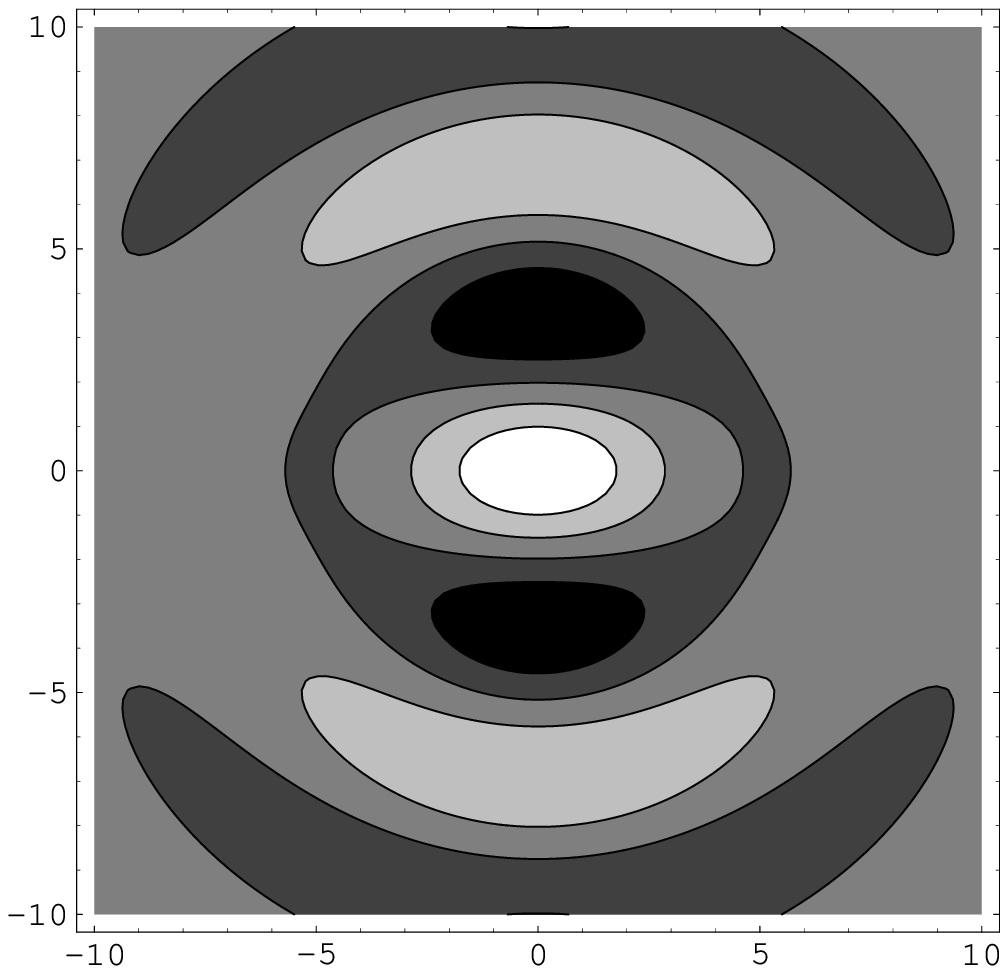}}
Figure 1
\end{figure}
\clearpage

\begin{figure}[p]
\resizebox{\textwidth}{!}{\includegraphics{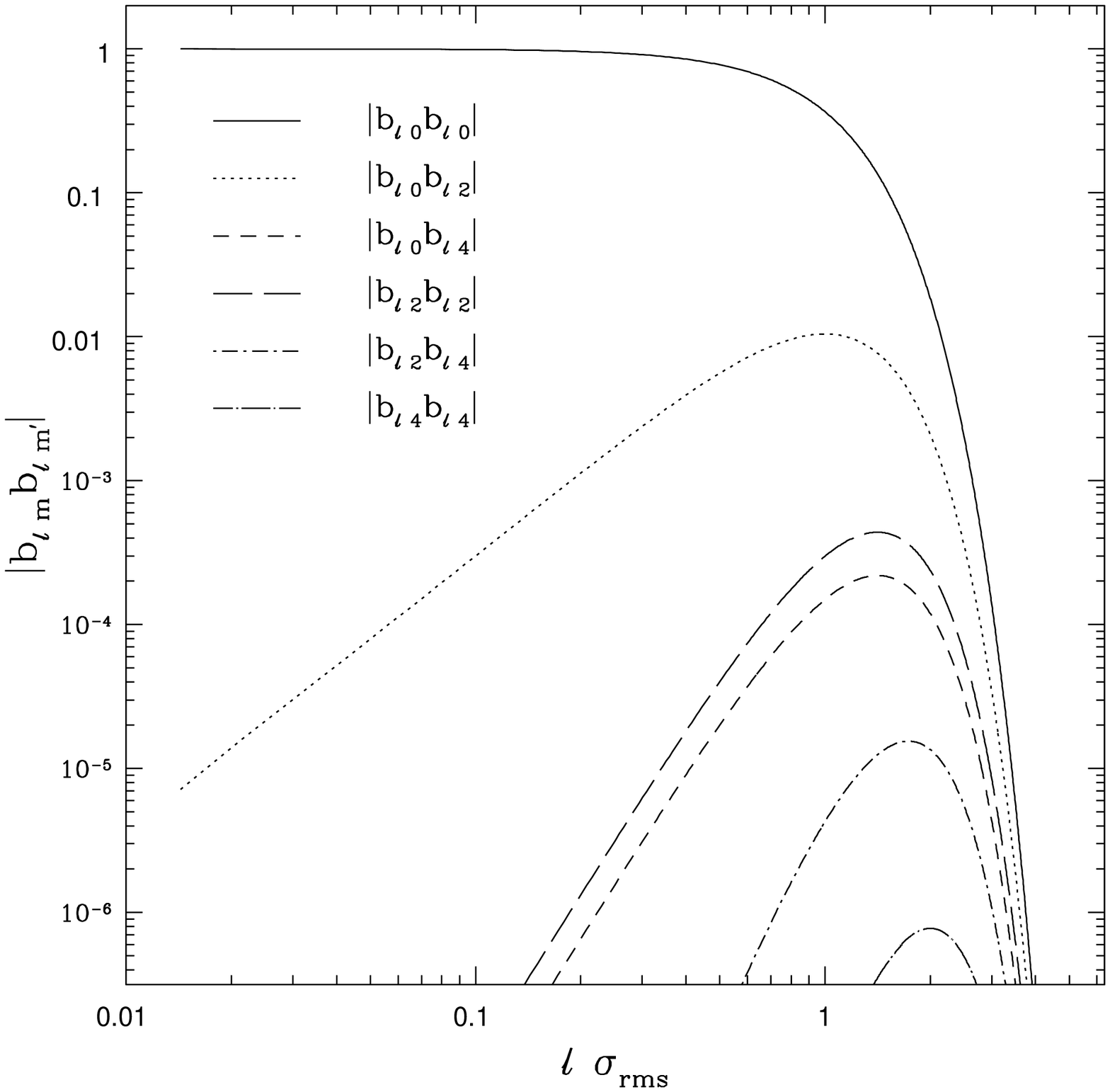}}
Figure 2
\end{figure}
\clearpage

\begin{figure}[p]
\resizebox{0.5\textwidth}{!}{\includegraphics{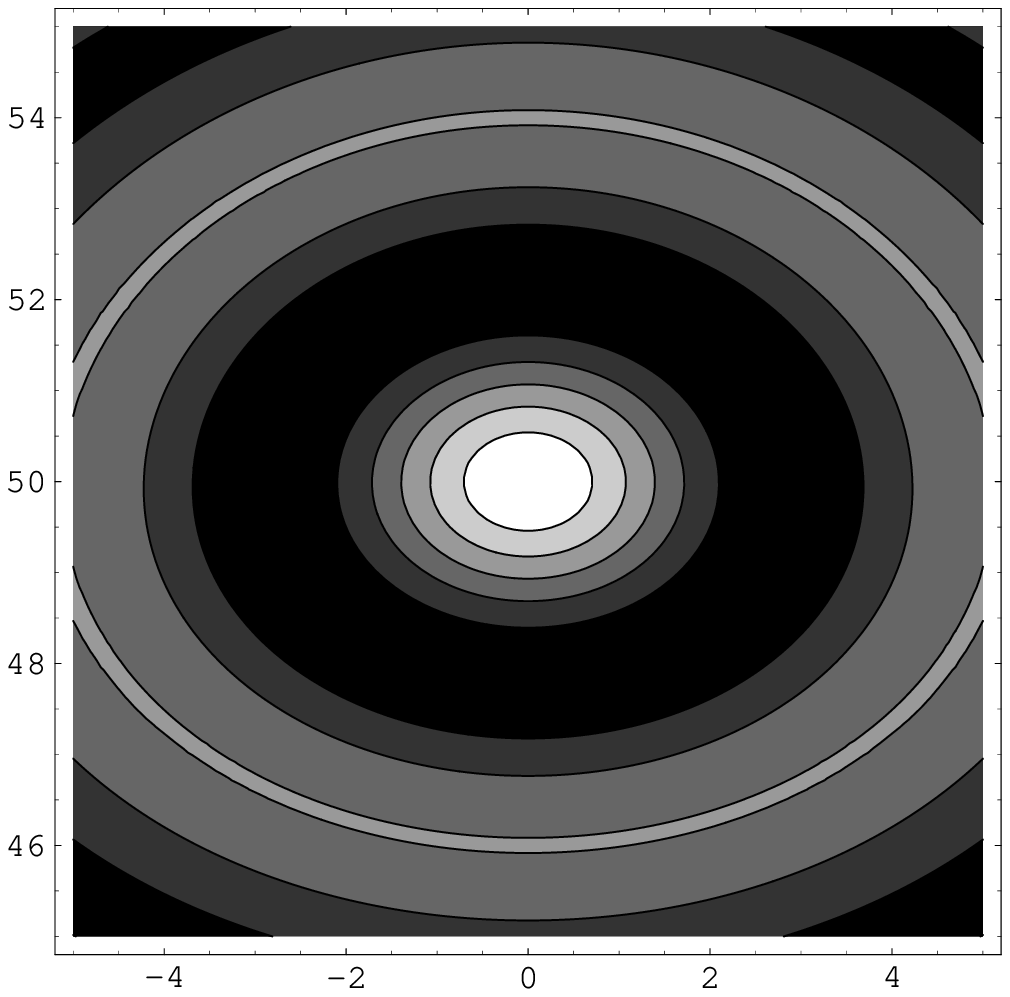}}
\resizebox{0.5\textwidth}{!}{\includegraphics{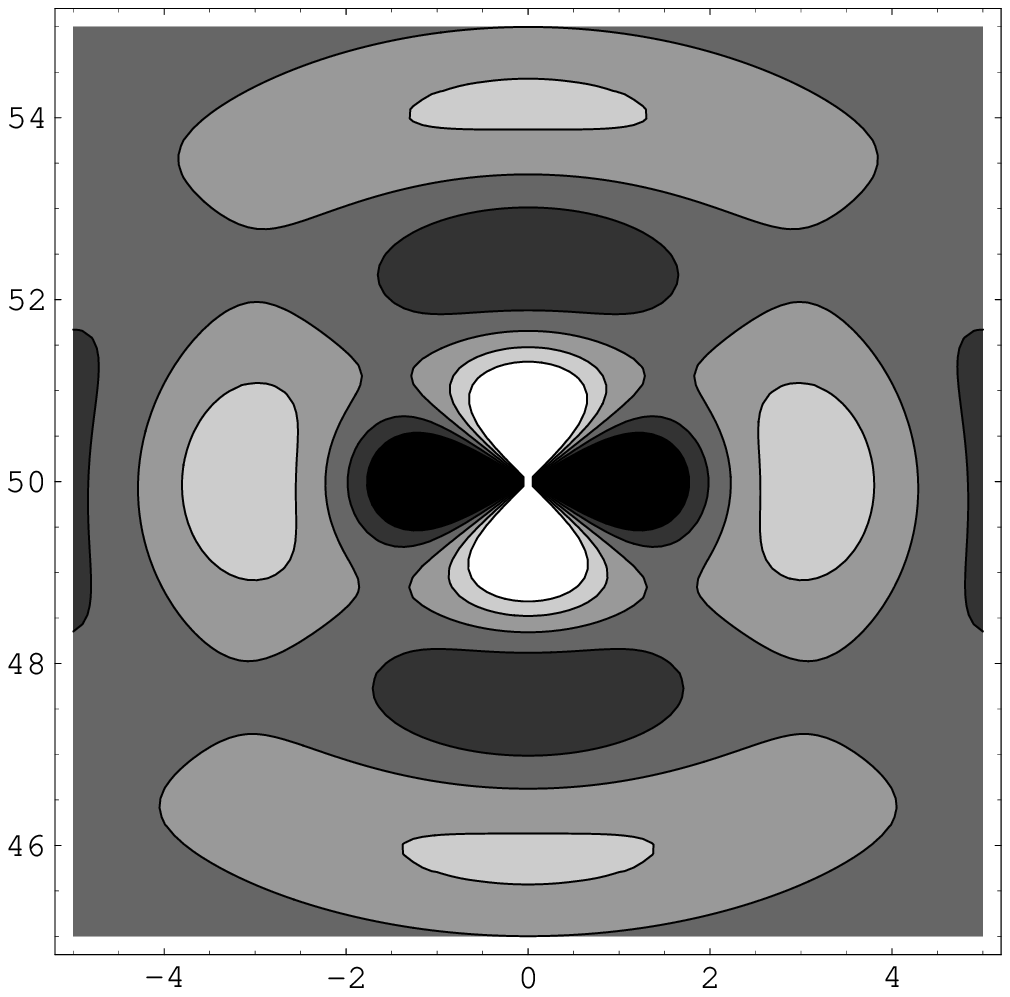}}
Figure 3
\end{figure}
\clearpage

\begin{figure}[p]
\resizebox{\textwidth}{!}{\includegraphics{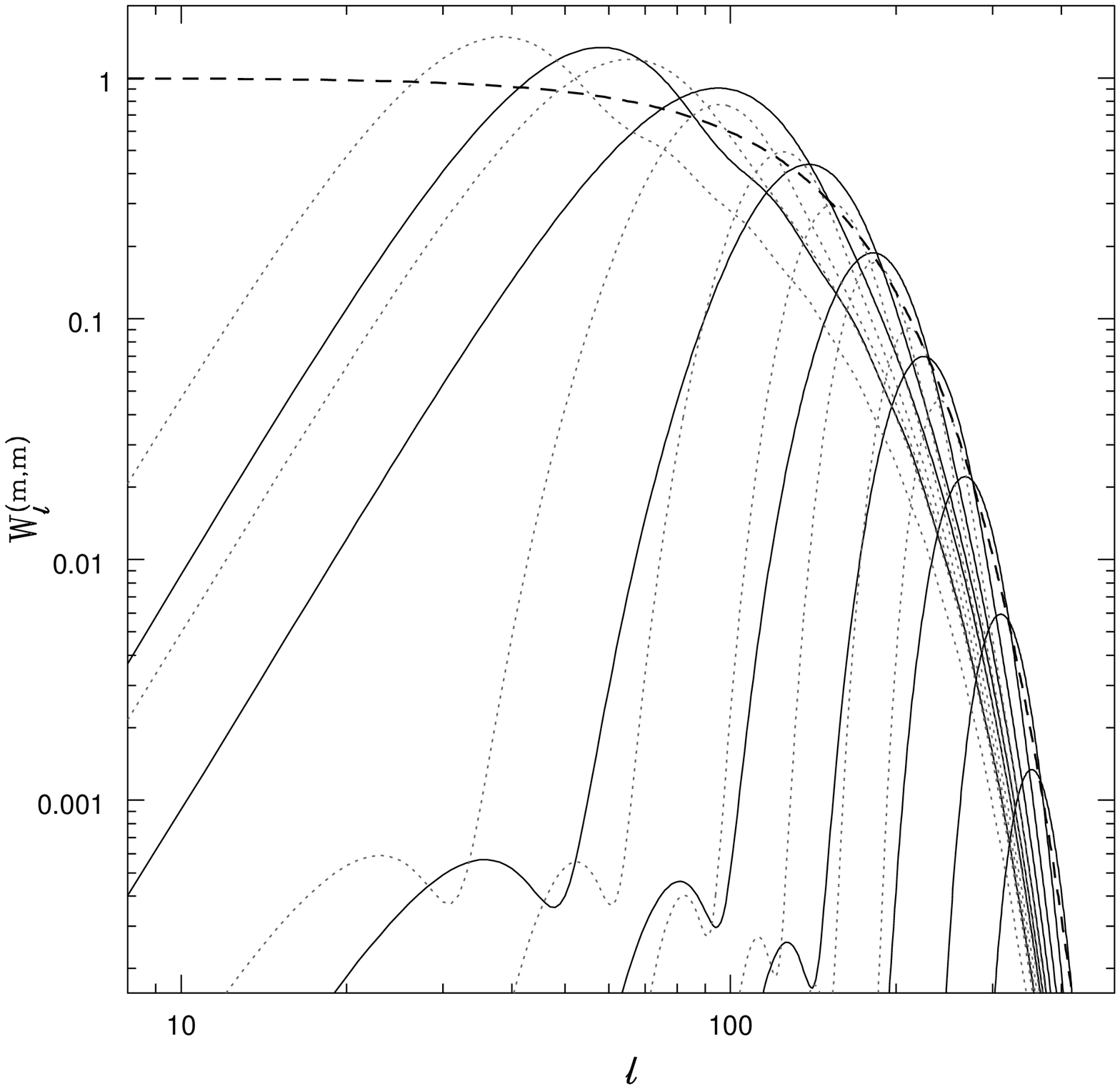}}
Figure 4
\end{figure}
\clearpage

\begin{figure}[p]
\resizebox{\textwidth}{!}{\includegraphics{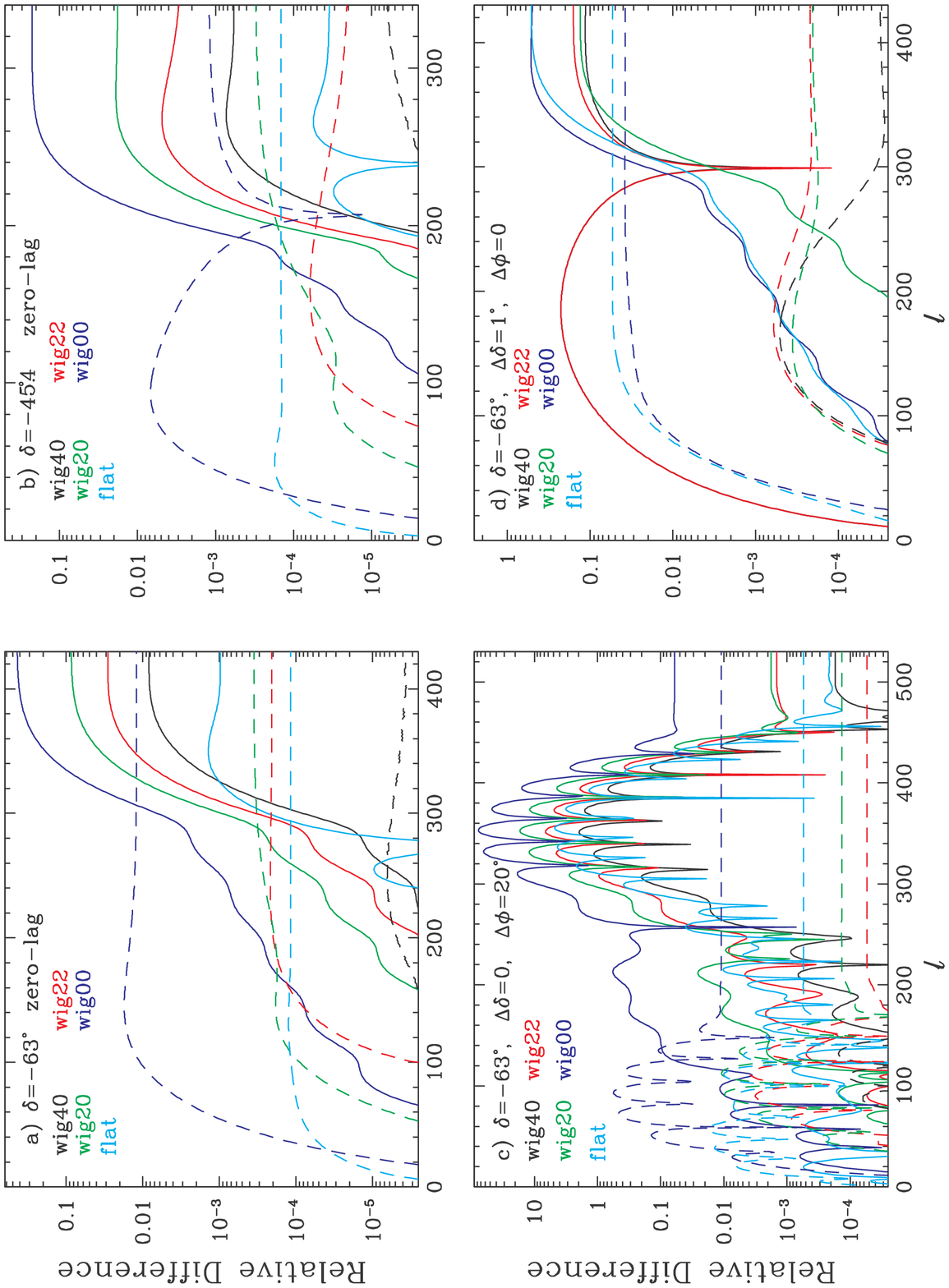}}
Figure 5
\end{figure}
\clearpage

\end{document}